\documentclass[aps,prd,preprint,nopreprintnumbers,nofootinbib,showpacs]{revtex4-1}

\usepackage{graphicx}
\usepackage{dcolumn}
\usepackage{bm}
\usepackage{epsfig}
\usepackage{amsmath}

\begin{document}

\newcommand{\ee}{e$^+$e$^-$}
\newcommand{\ff}{f$_{2}$(1525)}
\newcommand{\bb}{$b \overline{b}$}
\newcommand{\cc}{$c \overline{c}$}
\newcommand{\sbs}{$s \overline{s}$}
\newcommand{\uu}{$u \overline{u}$}
\newcommand{\dd}{$d \overline{d}$}
\newcommand{\qq}{$q \overline{q}$}
\newcommand{\suo}{\rm{\mbox{$\epsilon_{b}$}}}
\newcommand{\loro}{\rm{\mbox{$\epsilon_{c}$}}}
\newcommand{\kos}{\ifmmode \mathrm{K^{0}_{S}} \else K$^{0}_{\mathrm S} $ \fi}
\newcommand{\kol}{\ifmmode \mathrm{K^{0}_{L}} \else K$^{0}_{\mathrm L} $ \fi}
\newcommand{\ko}{\ifmmode {\mathrm K^{0}} \else K$^{0} $ \fi}

\def\tpc{three-particle correlation}
\def\twopc{two-particle correlation}
\def\ksks{K$^0_S$K$^0_S$}
\def\ee{e$^+$e$^-$}
\def\ff{f$_{2}$(1525)}

\title{Phase Difference Between the Electromagnetic and Strong Amplitudes for $\psi(2S)$ and $J/\psi$ Decays into Pairs of Pseudoscalar Mesons}

\author{Z. Metreveli}
\author{S. Dobbs}
\author{A. Tomaradze}
\author{T. Xiao}
\author{Kamal K. Seth}
\affiliation{Northwestern University, Evanston, Illinois 60208, USA}

\author{J. Yelton}
\affiliation{University of Florida, Gainesville, Florida 32611, USA}

\author{D. M. Asner}
\author{G. Tatishvili}
\affiliation{Pacific Northwest National Laboratory, Richland, Washington 99352, USA}

\author{G. Bonvicini}
\affiliation{Wayne State University, Detroit, Michigan 48202, USA}


\date{\today}

\begin{abstract}
Using the data for $24.5\times 10^{6}$ $\psi(2S)$ produced in $e^{+}e^{-}$ annihilations at 
$\sqrt{s}=3686$ MeV at the CESR-c $e^{+}e^{-}$ collider and $8.6\times10^6$ $J/\psi$ produced in the decay $\psi(2S)\to\pi^+\pi^-J/\psi$, the branching fractions for $\psi(2S)$ and $J/\psi$  
decays to pairs of pseudoscalar mesons, $\pi^{+}\pi^{-}$, $K^{+}K^{-}$, and $K_{S}K_{L}$,  
have been measured using the CLEO-c detector. We obtain  branching fractions  
$\mathcal{B}(\psi(2S)\to \pi^{+}\pi^{-})=(7.6\pm 2.5 \pm 0.6)\times 10^{-6}$, 
$\mathcal{B}(\psi(2S)\to K^{+}K^{-})=(74.8\pm 2.3\pm 3.9)\times 10^{-6}$,  
$\mathcal{B}(\psi(2S)\to K_{S}K_{L})=(52.8\pm 2.5\pm 3.4)\times 10^{-6}$,
and $\mathcal{B}(J/\psi\to \pi^{+}\pi^{-})=(1.47\pm 0.13 \pm 0.13)\times 10^{-4}$, $\mathcal{B}(J/\psi\to K^{+}K^{-})=(2.86\pm 0.09 \pm 0.19)\times 10^{-4}$,
 $\mathcal{B}(J/\psi\to K_{S}K_{L})=(2.62\pm 0.15 \pm 0.14)\times 10^{-4}$,
where the first errors are statistical and the second errors are systematic.
The phase differences between the amplitudes for electromagnetic and strong decays of 
$\psi(2S)$ and $J/\psi$ to $0^{-+}$ pseudoscalar pairs are determined by a Monte Carlo
method to be $\delta(\psi(2S)_{PP}=(110.5^{+16.0}_{-9.5})^{\circ}$ and  $\delta(J/\psi)_{PP}=(73.5^{+5.0}_{-4.5})^{\circ}$.  The difference between the two is $\Delta\delta \equiv \delta(\psi(2S))_{PP}-\delta(J/\psi)_{PP} =(37.0^{+16.5}_{-10.5})^\circ.$
\end{abstract}

\maketitle

\section{Introduction}

Interest in final state interaction (FSI) phases originally arose from CP violation
in $K$ decays and $B$ decays. However, recently interest has focused on the suggestion 
that large FSI phases are a general feature of strong decays. The electromagnetic and 
strong decays of the vector states of charmonium, $J/\psi$ and $\psi(2S)$, offer good
testing ground for this possibility. In a series of papers 
Suzuki~\cite{suzuki,suzuki1,suzuki2} has studied FSI phase differences between the 
electromagnetic amplitude $A_{\gamma}$ and the three-gluon strong amplitude $A_{ggg}$.
In the decays of $J/\psi$ to $1^{-}0^{-}$
vector-pseudoscalar (VP) pairs Suzuki obtained 
$\delta(J/\psi)_{VP}=80^{\circ}$~\cite{suzuki},
and for the $0^{-}0^{-}$ pseudoscalar-pseudoscalar (PP) pairs he obtained
$\delta(J/\psi)_{PP}=(89.6\pm 9.9)^{\circ}$~\cite{suzuki1}. 
Rosner~\cite{rosner} confirmed Suzuki's results, obtaining 
$\delta(J/\psi)_{VP}=(76^{+9}_{-10})^{\circ}$, and 
$\delta(J/\psi)_{PP}=(89\pm 10)^{\circ}$. 
Further, Gerard and Weyers~\cite{gerard} have argued that these differences are manifestations
of what they call ``universal incoherence'', i.e., $90^{\circ}$ phase difference between electromagnetic and 
every exclusive annihilation decay channel of $J/\psi$ and $\psi(2S)$.
In order to arrive at a deeper understanding 
of the origin of large phase differences it is therefore necessary to examine if what has been observed for $J/\psi$ decays persists in 
the corresponding decays of $\psi(2S)$. As Suzuki has noted~\cite{suzuki2}, this is
particularly important in the context of the curious suppression of the ratio 
$\Gamma[\psi(2S) \to VP]/\Gamma[J/\psi \to VP]$ (particularly notable for 
$\rho \pi$ decays).

One of the best places to study the phase difference between electromagnetic and strong
decays of $\psi(2S)$ and $J/\psi$ is in their decays to pseudoscalar pairs,
$\pi^{+}\pi^{-},~K^{+}K^{-}$, and $K_{S}K_{L}$. This is because the three
decays sample the interactions in quite different ways. The $\pi^{+}\pi^{-}$ decay is
essentially purely electromagnetic, with strong decay being forbidden by isospin invariance,
the $K_{S}K_{L}$ decay is essentially purely strong and due only to SU(3) violation,
and the $K^{+}K^{-}$ decay can proceed through both the electromagnetic and strong 
interactions.

A particularly simple and transparent determination
of the relative phase angle $\delta(\psi)$ can be made by measuring it as the 
{\it interior} angle of the triangle in the complex plane with the amplitudes for the 
decays to $\pi^{+}\pi^{-},~K^{+}K^{-}$, and $K_{S}K_{L}$ as its three sides. This is 
illustrated in Fig.~1 for the $J/\psi$ and $\psi(2S)$ decays.

With the neglect of the small effect of SU(3)--breaking and interference between
resonance and continuum amplitudes, the relative phase angle $\delta(\psi)$ is 
given by
\begin{equation} 
\delta(\psi)_{\mathrm{PP}}\!=\!\cos^{-1} \! \left( \frac{\mathcal{B}(K_{S}K_{L})\!+\!\rho\mathcal{B}(\pi^{+}\pi^{-})\!-\!\mathcal{B}(K^{+}K^{-}) }{ \vert 2\sqrt{\mathcal{B}(K_{S}K_{L}) \times \rho \times \mathcal{B}(\pi^{+}\pi^{-}) \vert }}  \right) , 
\end{equation} 
where the phase space correction factor $\rho =(p_{K}/p_{\pi})^{3}$; 
$\rho(\psi(2S))=0.902$, and $\rho(J/\psi)=0.862$. Thus it is required to measure the 
three branching fractions, 
$\mathcal{B}(\psi(2S),~J/\psi \to \pi^{+}\pi^{-},~K^{+}K^{-}$, and $K_{S}K_{L})$, which are 
proportional to the squares of the respective amplitudes. 

\begin{figure}[t!]       
\begin{center} 
\includegraphics[width=7.0cm]{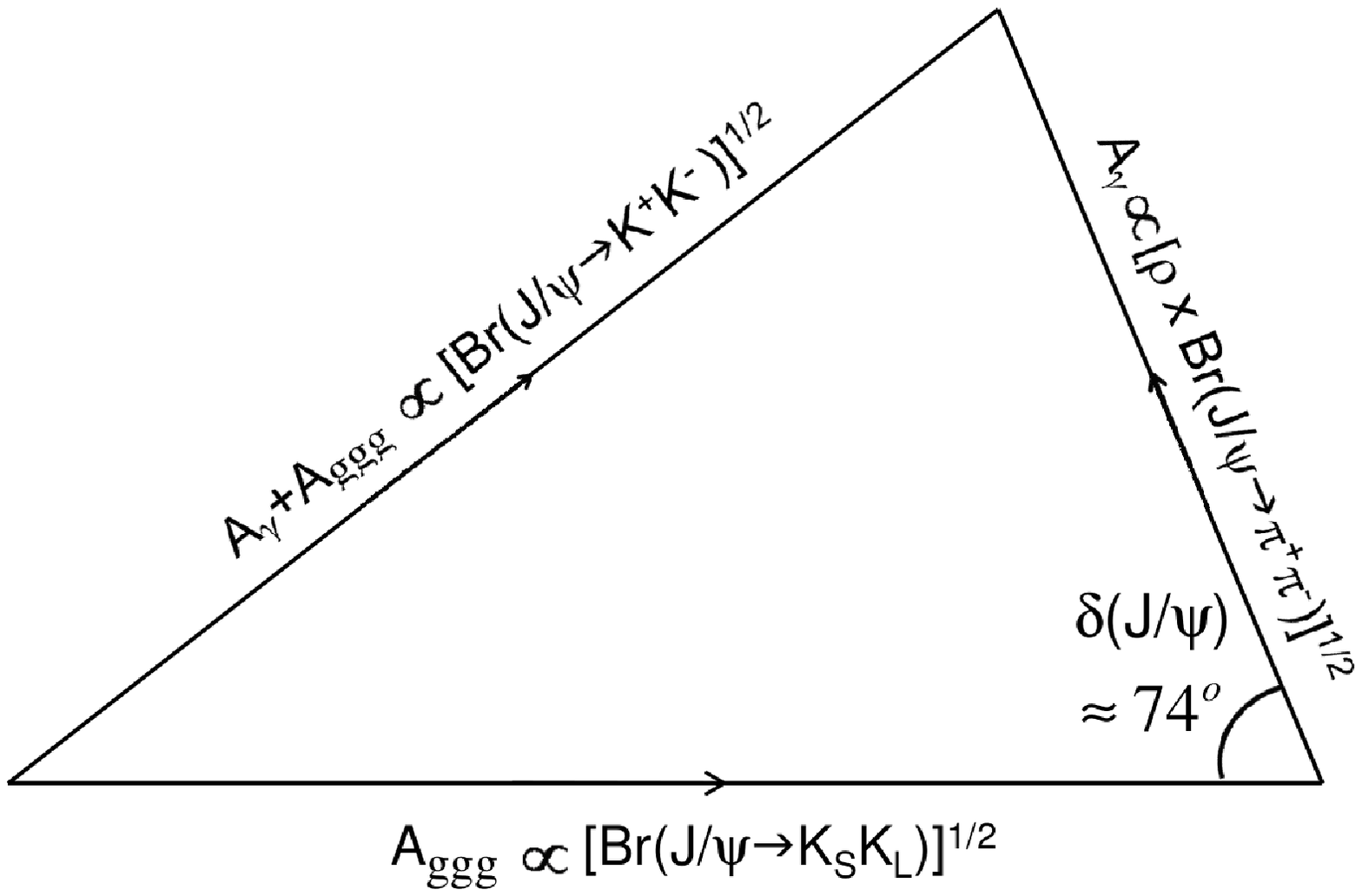}
\hspace*{0.3cm}      
\includegraphics[width=8.5cm]{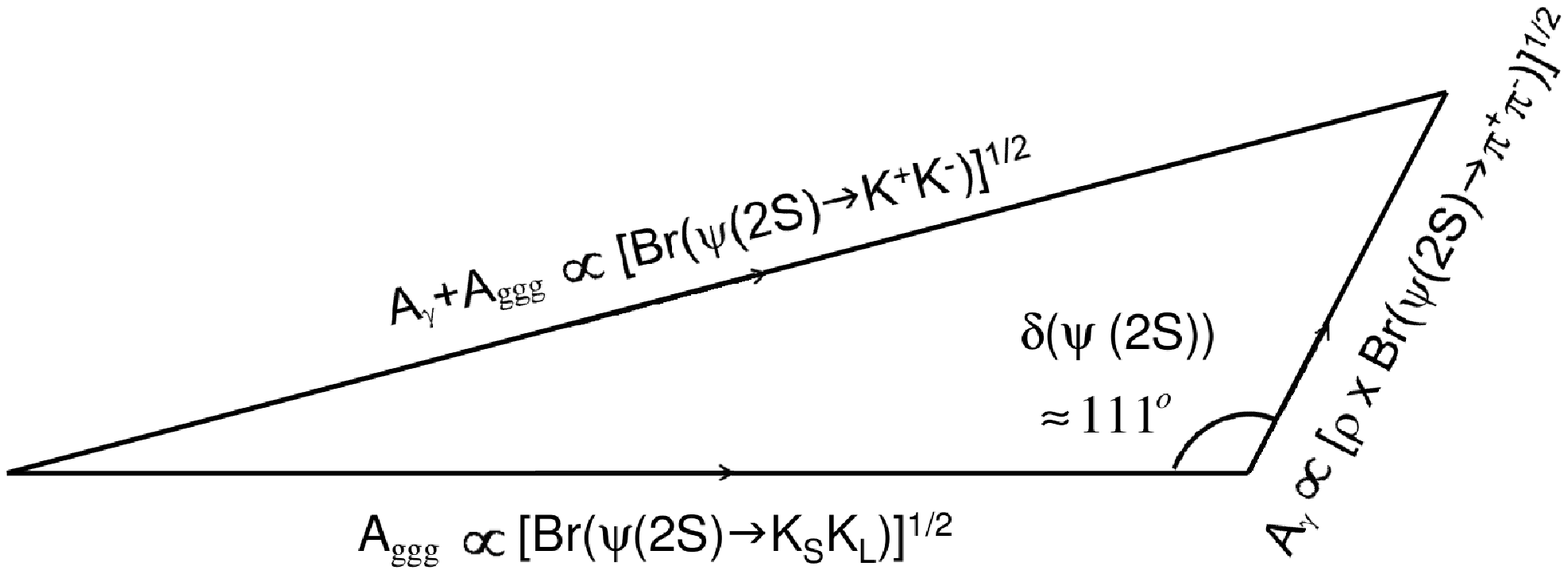} 
\end{center}       
\caption{
Triangle representations of the amplitudes in $J/\psi \to PP$ and $\psi(2S)\to PP$ decays. 
The relative phase angle between the strong and electromagnetic amplitudes is $\delta(\psi)$.
}        
\end{figure}       

Previous measurements of $\delta(\psi(2S))_{PP}$ were made by BES~\cite{bes1,bes} and
CLEO~\cite{cleo}. Their results, recalculated to correspond to the internal phase angle 
defined by Eq.~(1), are $\delta(\psi(2S))_{PP}=(91\pm35)^{\circ}$ (BES), and 
$\delta(\psi(2S))_{PP}=(87\pm20)^{\circ}$ (CLEO). The results for $\delta(J/\psi)_{PP}$
mentioned earlier were obtained by Suzuki and Rosner using the PDG 1998~\cite{pdg98}
evaluation of the branching fractions measured in 1985 by Mark~III~\cite{spear} in the decay of their sample
of $2.7\times 10^{6}~J/\psi$ produced directly in $e^{+}e^{-}$ annihilations at 
$\sqrt{s}=M(J/\psi)$.
In this paper we present much more accurate results for branching fractions and 
$\delta(\psi(2S))_{PP}$ using the CLEO-c data of $24.5\times 10^{6}$ $\psi(2S)$, eight
times larger than that used in the previous CLEO measurement~\cite{cleo}, and for branching 
fractions and $\delta(J/\psi)_{PP}$ using the data set of $8.6\times 10^{6}$ $J/\psi$
tagged by $\pi^{+}\pi^{-}$ recoils in the decay $\psi(2S)\to \pi^{+}\pi^{-} J/\psi$.
In all cases large improvement in precision over previous measurements is obtained.

The data used in this analysis were collected at the CESR $e^{+}e^{-}$ storage ring 
using the CLEO-c detector~\cite{cleodet}. The detector has a cylindrically symmetric 
configuration, and it provides 93$\%$ coverage of solid angle for charged and neutral 
particle identification. The detector components important for the present analysis 
are the vertex drift chamber, the main drift chamber (DR), the Ring Imaging Cherenkov 
(RICH) counter, and the CsI crystal calorimeter (CC).

\section{Event Selection}

Event selection, particle identification, and branching fraction determination in the
present paper follow closely those used by us in our published paper on the
determination of $\delta(\psi(2S))$~\cite{cleo}.

The events for the three decay modes, 
$\psi(2S),J/\psi \to \pi^{+}\pi^{-},K^{+}K^{-},K_{S}(\to\pi^{+}\pi^{-})K_{L}(\mathrm{not~detected})$ 
are required to have two charged particles for $\psi(2S)$ decays, and four charged particles for $J/\psi$ decays,
and zero net charge. The event selection criteria are identical to those in our earlier 
paper~\cite{cleo}. To recapitulate, all charged particles are required to meet the 
standard criteria for track quality. The $\pi^{+}\pi^{-}$,
$K^{+}K^{-}$, and $\pi^{+}\pi^{-}$ from $K_{S}$ decays (with vertex displaced by 
$> 5$~mm) are accepted in the regions $|\cos\theta|<0.75,~0.80$ and 0.93, respectively. 
The invariant mass of the $\pi^{+}\pi^{-}$ from $K_{S}$ decay is required to be within 
$\pm$10 MeV of $M(K_{S})=497.61$ MeV. 

Particle identification is done by combining $dE/dx$ information from the drift chamber (DR)
and the likelihood information from the RICH detector for the particle species
$i,~j\equiv p,~K,~\pi$, $\mu$, and $e$. The variable for the $dE/dx$ information is 
$S_{i}=[(dE/dx)_{\mathrm{measured}}-(dE/dx)_{\mathrm{expected}}]/\sigma(dE/dx)_{\mathrm{measured}}$, and for the RICH information it is the likelihood function, -2log$L$. To distinguish between two 
particle species a joint $\chi^{2}$ function, 
$\Delta \chi^{2}(i,j)= - 2 (\log L_{i} - \log L_{j}) + (S_{i}^{2}-S_{j}^{2})$ is 
constructed.
Charged kaons in $K^{+}K^{-}$ decays are distinguished from protons, pions, and leptons by
requiring $\Delta \chi^{2}(K,p/\pi/\mu/e)<-9$. Looser criteria are used for pions in
$\pi^{+}\pi^{-}$ decay and $\pi^{+}\pi^{-}$ daughters from $K_{S}$ in $K_{S}K_{L}$ decay,
$\Delta \chi^{2}(\pi,e/K/p)<0$. In addition, electrons are rejected in all decays by 
requiring $E(\mathrm{CC})/p<0.9$. All these requirements are identical to those in Ref.~\cite{cleo}.

In Ref.~\cite{zweber} a detailed study was made to distinguish pions from the much more
prolific yield of muons. It was determined that the energy deposited in the central
calorimeter by pions due to their hadronic interactions provides a very efficient means
of distinguishing them from muons which deposit much smaller energy due only to $dE/dx$.
For $\psi(2S)\to\pi^+\pi^-$ decay, it was determined that requiring every pion to deposit $E(\mathrm{CC})>0.42$ GeV reduced muon
contamination to $\ll 1\%$ level. This requirement was used in Ref.~\cite{cleo}, and we
impose it also in the present analysis for the channel $\psi(2S)\to \pi^{+}\pi^{-}$.
For $J/\psi\to \pi^{+}\pi^{-}$, the pion yield is much larger, and the corresponding requirement is determined to
be $E(\mathrm{CC})>0.35$ GeV.
For $K_{S}K_{L}$ decay an explicit $\pi^{0}$ veto was made, as in Ref.~\cite{cleo}.

\section{Determination of $\bm{\mathcal{B}(\psi(2S)\to \pi^{+}\pi^{-},~K^{+}K^{-}}$, $\bm{K_{S}K_{L})}$, and $\bm{\delta(\psi(2S))}$}

 For $\psi(2S)\to \pi^{+}\pi^{-}$ and $K^{+}K^{-}$, it was required that the
total momentum $|\Sigma p|$ be less than 75 MeV.  As expected, this requirement removes 
most of the $J/\psi$, $\chi_{cJ}$ 
peaks and the background which is present without it in the event distributions plotted as a function of $X(h)\equiv (E(h^+)+E(h^-)) / \sqrt{s}$.  The resulting $X(h)$ distributions are shown in Figs.~2(a,b) in the extended region of $X(h)$. For $\psi(2S)\to K_{S}K_{L}$ with only $K_{S}$ observed no such 
total momentum cut can be imposed, and the different criteria developed in 
Ref.~\cite{cleo} were implemented to take account of the unobserved $K_{L}$.

For $\psi(2S)\to K_{S}K_{L}$, the direction of $K_{L}$ is inferred from that of the observed
$K_{S}$. We require that there be no shower associated with neutrals closest to this 
direction with energy $> 1.5$ GeV.
Further, we require that in a cone of 0.35 radians around the $K_{L}$ direction there be
no more than one shower with $E_\mathrm{in}>100$ MeV. We require that outside this cone there be no 
single shower with $E_\mathrm{out}>100$ MeV and 
the sum of all showers $\Sigma E_\mathrm{out}<300$ MeV.
These selections remove events with neutral particles other than $K_{L}$ 
accompanying the detected $K_{S}$.  As shown in Fig.~2~(c), the remaining background is
featureless, and very small in the signal region, $X(K_S)\approx1.0$.

\begin{figure*}[tb!]      
\begin{center}      
\includegraphics[width=\textwidth]{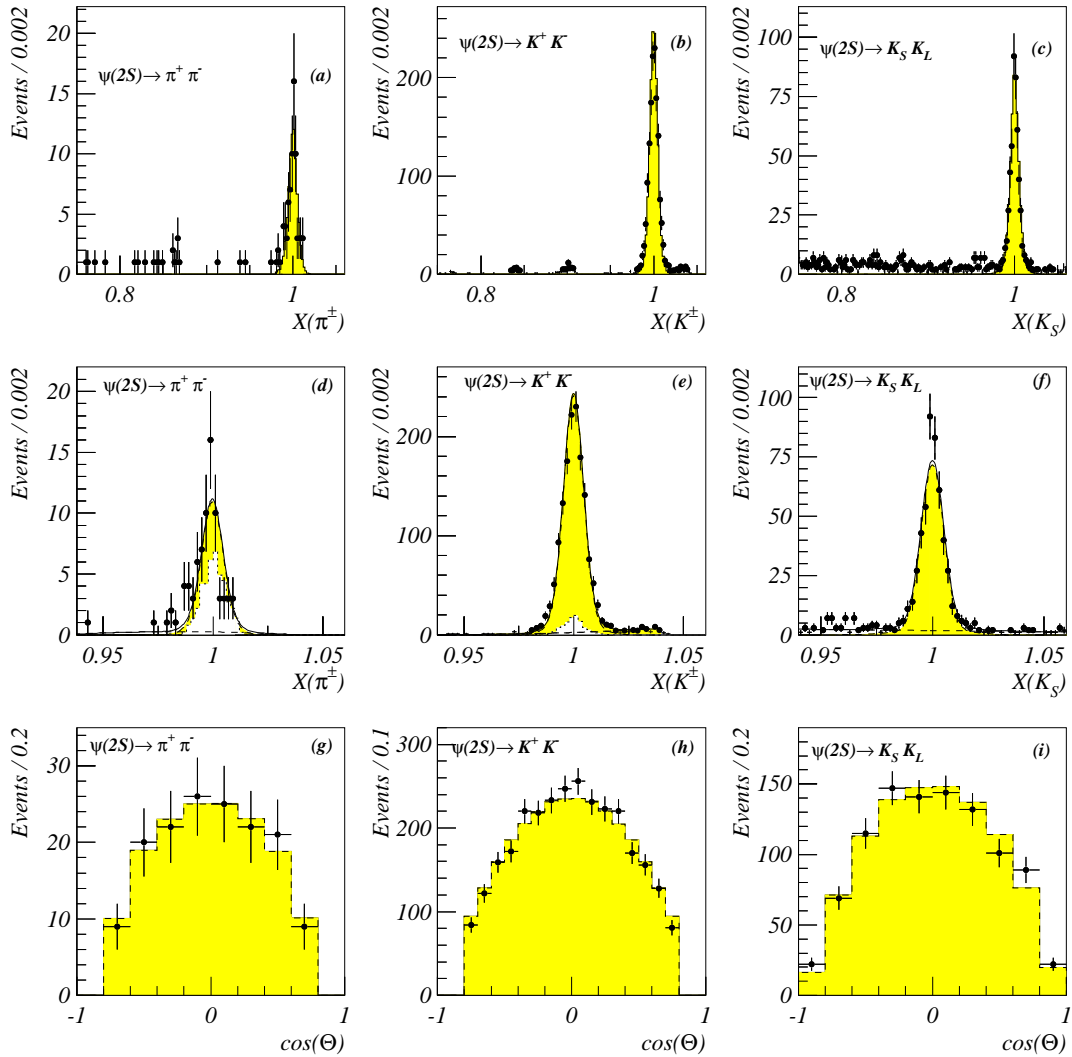} 
\end{center}      
\caption{Event distributions for $\psi(2S)\to\pi^+\pi^-$, $K^+K^-$, and $K_SK_L$.  Panels (a,b,c) show $X(h)$ distributions in the extended range $X(h)$. Panels (d,e,f) show enlarged $X(h)$ distributions in the region in which we fit the data with MC--determined peak shapes and linear backgrounds.  Panels (g,h,i), show the corresponding angular distributions.  Points with errors represent data; shaded histograms represent MC fits to the data.  In panels (d,e) the dotted histograms indicate continuum contributions.}       
\end{figure*}

In Figs.~2~(d,e,f) we show the $X(h)$ distributions of Figs.~2~(a,b,c) in detail in the 
smaller region of $X(h)$ in which we fit the data with MC--generated peak shapes and linear backgrounds.
The fit results are presented in Table~\ref{tbl:brresult}. The observed peak counts, 
$N(\mathrm{obs})$, are $70.8\pm 8.8$, $1431.3\pm 39.4$, and $478.0\pm 23.0$ for
$\pi^{+}\pi^{-}$, $K^{+}K^{-}$, and $K_{S}K_{L}$, respectively.
 
The MC signal events were generated assuming $\sin^{2}\theta$ angular distributions,
where $\theta$ is the angle between a meson and the positron beam.
In Figs.~2~(g,h,i) we show that the angular distributions of the data events are in excellent agreement 
with the MC distributions in all three cases for $\psi(2S)\to \pi^{+}\pi^{-},~K^{+}K^{-}$,
and $K_{S}K_{L}$ decays.

The observed peaks for $\pi^{+}\pi^{-}$ and $K^{+}K^{-}$ decays contain contributions
from continuum, or form factor production $e^{+}e^{-}\to \gamma^{*}\to \pi^{+}\pi^{-},
K^{+}K^{-}$ in addition to the resonance contributions. Continuum contribution in
$K_{S}K_{L}$ decays can, however, only arise from higher order processes, and is expected to
be very small.
The continuum contributions have to be estimated, and subtracted from the observed peak 
yields of $\pi^{+}\pi^{-}$ and $K^{+}K^{-}$ before the corresponding branching 
fractions can be determined. This is particularly challenging for the $\pi^{+}\pi^{-}$
decays. 

The main limitation in the measurement of the interference phase difference angle in  
earlier publications came from the determination of the branching fraction for 
$\pi^{+}\pi^{-}$ decay, $\mathcal{B}(\psi(2S)\to \pi^{+}\pi^{-})$. As mentioned earlier,
the three--gluon strong decay $\psi(2S)\to \pi^{+}\pi^{-}$ is forbidden by isospin 
conservation, and the branching fraction $\mathcal{B}(\psi(2S)\to \pi^{+}\pi^{-})$ is 
consequently small. 
The problem of the intrinsically small branching fraction is compounded by the fact that
no good--statistics measurements of the continuum $\pi^{+}\pi^{-}$ contribution were
available.
As a result all three previous measurements had very large $(60-100)\%$ errors: 
\begin{align} 
\nonumber \mathcal{B}(\pi^{+}\pi^{-})\times 10^{6} & = 80\pm50~(\mathrm{DASP~[15]}), \\
\nonumber & = 8.4\pm6.5~(\mathrm{BES~[6]}),~\mathrm{and}\\
\nonumber & = 8\pm8~(\mathrm{CLEO~[8]}).
\end{align} 
In the DASP~\cite{dasp} and BES~\cite{bes1} measurements no attempt was made to subtract the continuum contribution.
In our published paper~\cite{cleo} with 3 million $\psi(2S)$ (corresponding to 
integrated luminosity
$\int \mathcal{L} dt=5.6~\mathrm{pb}^{-1}$), only 11 $\pi^{+}\pi^{-}$ counts were observed.
From these the scaled continuum contribution of 7 counts, based on data with $e^{+}e^{-}$
integrated luminosity of 20.7 $\mathrm{pb}^{-1}$ taken off--$\psi(2S)$ at 
$\sqrt{s}=3670$ MeV, was subtracted to get $N(\pi^{+}\pi^{-})=4\pm 4$. 
This led to the poor determination of 
$\mathcal{B}(\psi(2S)\to \pi^{+}\pi^{-})=(8\pm8\pm2)\times 10^{-6}$~\cite{cleo}.

Our present analysis has two big advantages over the old analysis. We now have available a 
much larger data set of 24.5 million $\psi(2S)$ corresponding to an integrated 
$e^{+}e^{-}$ luminosity 
of 48 $\mathrm{pb}^{-1}$. Also, we are able to make a much better estimate of the continuum 
contributions in the yields of 
$\pi^{+}\pi^{-}$ and $K^{+}K^{-}$ based on our large-statistics form factor measurements 
with luminosity of 805 $\mathrm{pb}^{-1}$ at $\sqrt{s}=3772$ MeV and 586 pb$^{-1}$ at
$\sqrt{s}=4170$ MeV. 

The widths of $\psi(3770)$ and $\psi(4160)$ are $\Gamma(\psi(3770))=(27.3\pm 1.0)$ MeV and
$\Gamma(\psi(4160))=(103\pm 8)$ MeV, respectively~\cite{pdg10}. They are  about two 
orders of magnitude or more larger than $\Gamma(\psi(2S))=(0.304\pm 0.009)$ MeV, and the  
estimates of the resonance branching fractions for the decays of $\psi(3770)$ and 
$\psi(4160)$ to $\pi^{+}\pi^{-}$, $K^{+}K^{-}$, and $K_{S}K_{L}$ range from $1\times 10^{-8}$ to $8\times 10^{-8}$. 
These lead to the conclusion that the resonance contributions in $\pi^+\pi^-$ and $K^+K^-$ decays of $\psi(3770)$ and $\psi(4160)$ are less than 0.01\% of the total, 
i.e., the total observed counts $N'(\mathrm{tot.~obs})$  are entirely due to continuum
or form factor contribution. They can therefore be confidently extrapolated to estimate
continuum contribution in the measured counts at $\psi(2S)$. The extrapolation is done as
\begin{equation}
C\equiv \frac{N(\mathrm{cont},\sqrt{s})}{N'(\mathrm{obs},\sqrt{s'})} = {\mathcal{L}(\sqrt{s})\epsilon(\sqrt{s}) \over \mathcal{L}(\sqrt{s^{\prime}})\epsilon(\sqrt{s^{\prime}})} \times \left({\sqrt{s^{\prime}} \over \sqrt{s}} \right)^{6}.
\end{equation}
where $\sqrt{s}=M(\psi(2S)) = 3686$~MeV and $\sqrt{s^{\prime}}=3772$~MeV and 4160~MeV.
Here $\mathcal{L}(\sqrt{s},\sqrt{s'})$ are the luminosities, and $\epsilon(\sqrt{s},\sqrt{s'})$ are efficiencies determined by Monte Carlo (MC) simulations,  and the factor $(\sqrt{s^{\prime}}/\sqrt{s})^{6}$ is the conventional extrapolation based on the
constancy of $Q^{2}F(Q^{2})$ for vector meson form factors~\cite{explain}.
The observed $K^{+}K^{-}$ and $\pi^{+}\pi^{-}$ counts at $\sqrt{s^{\prime}}=3772$ MeV
and 4170 MeV have statistical and systematic uncertainties~\cite{zweber}.
Using the counts at 3772 and 4170 MeV in Eq.~(2) leads to estimated continuum
contributions at $\sqrt{s}=3686$ MeV which are 105.7$\pm$3.6$\pm$4.7 and 
109.2$\pm$4.9$\pm$4.8 counts
respectively for kaons, and 41.8$\pm$2.2$\pm$4.3 and 37.6$\pm$3.1$\pm$3.9 counts 
respectively for pions (the first errors are statistical and the second errors are
systematic). 
We use their averages, taking account of the fact that systematic errors are 
correlated, as 106.9$\pm$5.5 counts for kaons, and 40.4$\pm$4.6 counts for pions as 
our best estimates of continuum contributions at 3686 MeV.

These contributions are illustrated as dotted histograms in Figs.~2~(d,e).

In Table~\ref{tbl:brresult}, for $\psi(2S)$ decays we list the number $N(\mathrm{obs})$ of counts observed in the 
$\pi^{+}\pi^{-}$, $K^{+}K^{-}$ and $K_{S}K_{L}$ peaks,  the number $N(\mathrm{cont})$ of continuum counts estimated as described above, the net signal counts $N(\mathrm{signal})=N(\mathrm{obs})-N(\mathrm{cont})$, the event selection efficiencies $\epsilon$, and the branching fractions calculated as
\begin{equation}
\mathcal{B}(\psi(2S)\to PP)=N(\mathrm{signal})/[\epsilon\times N(\psi(2S))],
\end{equation}  
where the number of $\psi(2S)$ is $N(\psi(2S))=24.5\times 10^{6}$.
We note that these branching fractions have been obtained without taking account
of possible interference between continuum and resonance contributions.

We have considered various sources of systematic uncertainties in our branching fraction
results. As in Ref.~\cite{cleo}, for all three decay channels, 
$\psi(2S)\to \pi^{+}\pi^{-},~K^{+}K^{-}$, and
$K_{S}K_{L}$ the common uncertainties are $\pm 2\%$ in number of $\psi(2S)$,
$\pm 1\%$ per track in track finding, and $\pm 1\%$ per track in charged
particle identification. Uncertainties in trigger efficiency are $\pm 1\%$
in $\pi^{+}\pi^{-}$ and $K^{+}K^{-}$, and $\pm 2\%$ in $K_{S}K_{L}$.
The systematic uncertainty in determination of the factor $C$ in Eq.~(2) comes from the
uncertainties in the total integrated luminosity values of the data taken at 
$\psi(2S)$ and at $\sqrt{s}=3772$ MeV and $\sqrt{s}=4170$ MeV, which correspond to 
1$\%$ for each. Thus, we assign 1.4$\%$ systematic uncertainty to the value $C$, 
determined in Eq.~(2). 

Variation of the total momentum $\Sigma p_{i}<$~75~MeV requirement by $\pm 15$ MeV resulted 
in no statistically significant change in $\mathcal{B}(\psi(2S)\to \pi^{+}\pi^{-})$, and
a $\pm 3.5\%$ change in $\mathcal{B}(\psi(2S)\to K^{+}K^{-})$, which we assign as systematic
uncertainty. For $\psi(2S)\to \pi^{+}\pi^{-}$ changing the requirement $E(CC)>0.42$ GeV
by $\pm 10\%$ resulted in $\pm 7\%$ change in $\mathcal{B}(\psi(2S)\to \pi^{+}\pi^{-})$.
The effect of the implementation of $K_{L}$-related constraints in $K_{S}K_{L}$ decay was
determined as the ratio of fitted peak counts in the spectra in Fig.~2(f)/Fig.~2(c).
It was determined to be 0.835$\pm$0.042. The 5$\%$ uncertainty in this determination was
assigned as a systematic error in $\mathcal{B}(\psi(2S)\to K_{S}K_{L})$. 

The total systematic uncertainties are 8.0$\%$, 5.2$\%$, and 6.5$\%$ for 
$\psi(2S)\to \pi^{+}\pi^{-},~K^{+}K^{-}$, and $K_{S}K_{L}$ decays, respectively.

The resulting branching fractions are
\begin{eqnarray}
& \nonumber\mathcal{B}(\psi(2S)\to \pi^{+}\pi^{-})=[7.6\pm2.5(\mathrm{stat})\pm 0.6(\mathrm{syst})]\times 10^{-6}, & \\
& \nonumber\mathcal{B}(\psi(2S)\to K^{+}K^{-})=[7.48\pm0.23(\mathrm{stat})\pm 0.39(\mathrm{syst})]\times 10^{-5}, & \\
& \mathcal{B}(\psi(2S)\to K_{S}K_{L})=[5.28\pm0.25(\mathrm{stat})\pm 0.34(\mathrm{syst})]\times 10^{-5}. &
\end{eqnarray}
These are listed in Table~\ref{tbl:brresult}. In Table~\ref{tbl:deltaresult} the errors are listed with the statistical and 
systematic errors combined in
quadrature, together with results from previous investigations 
by DASP~\cite{dasp}, BES~\cite{bes,bes2}, and CLEO~\cite{cleo}. 
The uncertainties in our branching fractions results are factors two to five smaller 
than those in the published results. In Table~\ref{tbl:deltaresult}, we also list the phase angle difference $\delta(\psi(2S))=(113.4\pm11.5)^{\circ}$, calculated using Eq.~(1). 
All the published values of $\delta(\psi(2S))$ are also listed, as recalculated using Eq.~(1).

In Sec.~V we present the determination of the phase angle differences using a MC method which allows us to take account of the distributions in the values of the branching fractions.

\section{Determination of $\bm{\mathcal{B}(J/\psi\to \pi^{+}\pi^{-},~K^{+}K^{-},~K_{S}K_{L})}$ and $\bm{\delta(J/\psi)}$}

The 24.5 million $\psi(2S)$ in our data set lead to 8.6 million $J/\psi$ events tagged by
$\pi^{+}\pi^{-}$ recoil in the decay $\psi(2S)\to \pi^{+}\pi^{-} J/\psi$. This sample
is automatically free of any contamination of $e^{+}e^{-}$, $\mu^{+}\mu^{-}$,
$K^{+}K^{-}$, or $p\bar{p}$ events produced in direct measurement at $\sqrt{s}=3097$ MeV. 
The subsequent $J/\psi$
decays to $\pi^{+}\pi^{-}$, and $K^{+}K^{-}$ also do not contain any continuum contributions.
As such, these data are cleaner and simpler to analyze than the data from $e^{+}e^{-}$
collisions at $\sqrt{s}=3097$ MeV.

As stated earlier, to select events for $J/\psi$ decays our event selection criteria are
modified to require four charged particles instead of two. Recoil mass is then constructed
for every pair of two oppositely charged particles. This is dominated by the production 
of $J/\psi$, as shown in Fig.~3 for the channel 
$J/\psi \to K_{S}K_{L}$, which is similar to those obtained for the channels 
$J/\psi \to \pi^{+}\pi^{-},~K^{+}K^{-}$. We define the clean $J/\psi$ sample as consisting of events with
$M(\mathrm{recoil})=M(J/\psi)\pm 10$ MeV. Selection of events for 
$J/\psi\to\pi^{+}\pi^{-},~K^{+}K^{-}$, and $K_{S}K_{L}$ is done exactly in the same manner
as described for $\psi(2S)$. 
As mentioned earlier, the most efficient $E(CC)$ cut to
reject $\mu^{+}\mu^{-}$ from $J/\psi$ decays is to require that each pion satisfy
$E(CC)>350$ MeV. The MC--estimated muon contamination with this requirement is $<1\%$.

\begin{figure}[tb!]         
\begin{center}         
\includegraphics[width=0.5\textwidth]{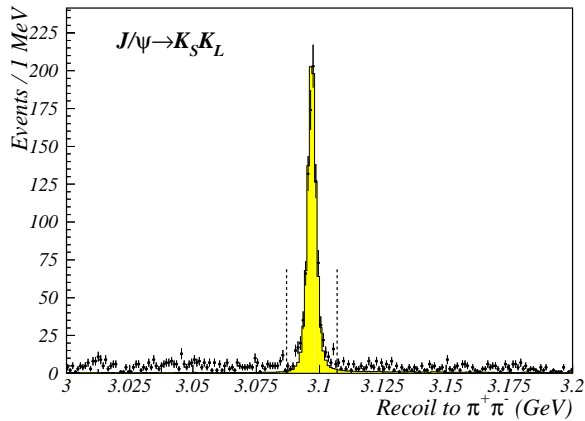}   
\end{center}         
\caption{ For the decay $\psi(2S)\to \pi^{+}\pi^{-}J/\psi$, $J/\psi\to K_{S}K_{L}$, $K_{S}\to \pi^{+}\pi^{-}$,
distributions of the recoil invariant mass against $\pi^{+}\pi^{-}$ pair which does not come 
from the $K_{S}$ decay.  Points are for the data, shaded histogram corresponds to the
signal MC. Normalization is arbitrary. }          
\end{figure}        

\begin{figure*}[tb!]       
\begin{center}       
\includegraphics[width=\textwidth]{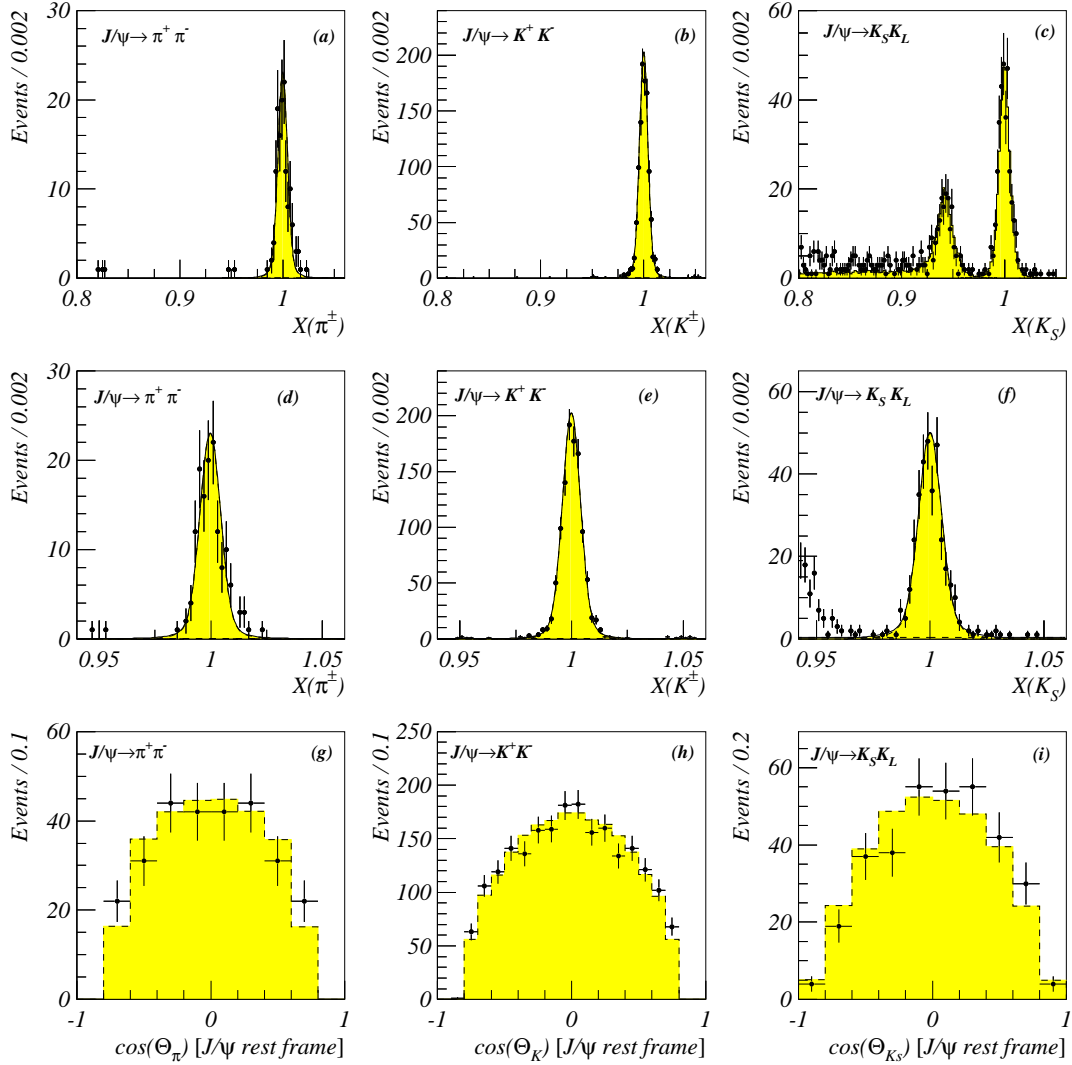}    
\end{center}       
\caption{Event distributions for $J/\psi\to\pi^+\pi^-$, $K^+K^-$, and $K_SK_L$.  
Panels (a,b,c) show $X(h)$ distributions in the extended range $X(h)=0.8-1.1$. 
Panels (d,e,f) show enlarged $X(h)$ distributions in the region $X(h)=0.94-1.06$ 
with fits with MC determined peak shape and linear background. Panels~(g,h,i) show 
the corresponding angular distributions.  Points with errors represent data, 
shaded histograms represent MC fits to the data.}        
\end{figure*}   
 
Figures~4~(g,h,i), show that the angular distributions of the data events for all thr\
ee 
$J/\psi$ decays are in excellent agreement with the MC $\sin^2\theta$ distributions, 
as in the case of $\psi(2S)$ decays in Figs.~2~(g,h,i). 
 
The systematic errors for $J/\psi$ decays were determined in exactly the same manner \
as  
for $\psi(2S)$ decays. 
Their totals are $9\%$, $6.8\%$, and $5.5\%$ for $\pi^{+}\pi^{-}$, $K^{+}K^{-}$, and 
$K_{S}K_{L}$, respectively.  

Figure~4 shows the results for $J/\psi$ decays to $\pi^+\pi^-$, $K^+K^-$, $K_SK_L$ as Fig.~2 does for the corresponding $\psi(2S)$ decays.  Figures~4~(a,b,c)  show the data in the extended range of $X(h)$ with arbitrarily normalized MC predictions.  The distribution of
 the $K_{S}$ events as a function of $X(K_{S})$ in 
the rest frame of $J/\psi$, shown in Fig.~4~(c), needs comment.
 The $K_{S}K_{L}$ peak at $X(K_{S})=1.0$ is clearly separated from the smaller
peak at $X\approx0.94$ which arises from the $J/\psi$ decays into $\overline{K}^{*0}(892)K_S$, $K^{*0}\to K_L\pi^0$, $K_S\to\pi^+\pi^-$, despite $\pi^0$ rejection. 
The clear separation of the $K_SK_L$ peak from the $\overline{K}^0(892)K_S$ peak is confirmed by the MC simulation whose result is superposed on the data in Fig.~4~(c).
In Figs.~4~(d,e,f), the event distributions of Figs.~(a,b,c) are shown in detail in the smaller region of $X(h)$.  Also shown are the fits using MC--determined peak shapes and linear backgrounds.
The fits give
$137.6\pm 11.8$, $1057.0\pm 32.8$, and $334.3\pm 19.3$ counts for $\pi^{+}\pi^{-}$,
$K^{+}K^{-}$, and $K_{S}K_{L}$, respectively. These compare with 84, 107 and 74 counts in the Mark~III measurements with a factor three smaller sample of $J/\psi$~\cite{spear}. These counts, the MC determined efficiencies $\epsilon$, and
$N(J/\psi)=(8.57\pm 0.07)\times 10^{6}$~\cite{njpsi}, lead to the branching 
fractions listed in Table~\ref{tbl:brresult}.

\begin{table*}[tb!]    
\begin{center}    
\begin{tabular}{lcccccc}    
\hline  \hline
 & $N(\mathrm{obs})$ & $N(\mathrm{cont})$ & $N(\mathrm{signal})$ & $\epsilon$ (\%) & $\mathcal{B}(\psi(2S)\to PP)\times 10^{6}$ & $\chi^{2}/\mathrm{dof}$ \\ 
\hline
$\psi(2S)\to \pi^{+}\pi^{-}$ & $70.8\pm 8.8$ & $40.4\pm4.6$ & $30.4\pm9.9$ & 16.4 & {\bf 7.6$\pm$2.5$\pm$0.6} & 0.68 \\ 
$\psi(2S)\to K^{+}K^{-}$ & $1431.3\pm39.4$ & $106.9\pm5.5$ & $1324.4\pm39.8$ & 72.4 & {\bf 74.8$\pm$2.3$\pm$3.9} & 1.11 \\ 
$\psi(2S)\to K_{S}K_{L}$ & $478.0\pm23.0$ & & $478.0\pm23.0$ &  53.5 & {\bf 52.8$\pm$2.5$\pm$3.4} & 1.00 \\ 
\hline 
 & & & & & $\mathcal{B}(J/\psi)\to PP)\times 10^{4}$ & \\ 
\hline
$J/\psi\to \pi^{+}\pi^{-}$ & $137.6\pm 11.8$ & & $137.6\pm 11.8$ & 10.9 & {\bf 1.47$\pm$0.13$\pm$0.13} & 1.09 \\ 
$J/\psi\to K^{+}K^{-}$ & $1057.7\pm 32.8$ & & $1057.7\pm 32.8$ & 43.1 & {\bf 2.86$\pm$0.09$\pm$0.19} & 1.00 \\ 
$J/\psi\to K_{S}K_{L}$ & $334.3\pm 19.3$ & & $334.3\pm 19.3$ & 21.5 & {\bf 2.62$\pm$0.15$\pm$0.14} & 1.03 \\ 
\hline \hline

\end{tabular}    
\end{center}    
\caption{
Fit results for $\psi(2S),~J/\psi \to \pi^{+}\pi^{-},~K^{+}K^{-}$, and $K_{S}K_{L}$ decays,
and the corresponding branching fractions. 
}    
\label{tbl:brresult}
\end{table*}

\begin{table*}[tb!]     
\begin{center}     
\begin{tabular}{lccccc}     
\hline \hline 
 & DASP~\cite{dasp} & BES~\cite{bes1,bes} & CLEO~\cite{cleo} & This analysis & This analysis \\ 
 & 1979 & 2004 & 2005 & & MC result \\ 
\hline
$\mathcal{B}(\psi(2S)\to\pi^{+}\pi^{-})\times 10^{6}$ & $80\pm 50$ & $8.4\pm6.5$ & $8\pm8$ & $\bm{7.6\pm2.6}$ & \\ 
$\mathcal{B}(\psi(2S)\to K^{+}K^{-})\times 10^{6}$ & $100\pm70$ & $61\pm21$ & $63\pm7$ & $\bm{74.8\pm4.5}$ & \\ 
$\mathcal{B}(\psi(2S)\to K_{S}K_{L})\times 10^{6}$ & --- & $52.4\pm6.7$ & $58.0\pm9.0$ & $\bm{52.8\pm4.2}$ & \\ 
$\delta(\psi(2S))$ & --- & $(91\pm35)^{\circ}$ & $(87\pm20)^{\circ}$ & $\bm{(113.4\pm 11.5)^{\circ}}$ & $\bm{(110.5^{+16.0}_{-9.5})^{\circ}}$ \\ 
\hline 
\end{tabular}
\begin{tabular}{@{\extracolsep{15.pt}}cccccc}
& Mark III\cite{spear} & BES~\cite{bes2} & This analysis & This analysis \\ 
& & & & MC result \\
\hline
$\mathcal{B}(J/\psi\to\pi^{+}\pi^{-})\times 10^{4}$ & $1.58\pm 0.25$ & --- & $\bm{1.47\pm0.18}$ & \\ 
$\mathcal{B}(J/\psi\to K^{+}K^{-})\times 10^{4}$ & $2.39\pm 0.33$ & --- & $\bm{2.86\pm0.21}$ & \\ 
$\mathcal{B}(J/\psi\to K_{S}K_{L})\times 10^{4}$ & $1.01\pm 0.18$ & $1.82\pm 0.13$ & $\bm{2.62\pm0.21}$ & \\ 
$\delta(J/\psi)$ & $(88\pm 11)^{\circ}$ & --- & $\bm{(73.6\pm5.6)^{\circ}}$ & $\bm{(73.5^{+5.0}_{-4.5})^{\circ}}$ \\ 
\hline \hline 

\end{tabular}     
\end{center}     
\caption{
Summary of results for $\psi(2S)$ and $J/\psi$ decays to pseudoscalar pairs:
branching fractions and the phase difference angles $\delta(\psi(2S))$ and
$\delta(J/\psi)$ using central values of branching fractions. 
The BES~\cite{bes} and CLEO~\cite{cleo} results for $\delta(\psi(2S))$
have been recalculated to correspond to the internal angle of the amplitude 
triangle. In the last column $\delta(\psi)$ results based on Monte Carlo calculation
described in the text are presented.
}     
\label{tbl:deltaresult}
\end{table*}

Thus the final branching fractions are
\begin{eqnarray}
& \nonumber\mathcal{B}(J/\psi\to \pi^{+}\pi^{-})=[1.47\pm0.13(\mathrm{stat})\pm 0.13(\mathrm{syst})]\times 10^{-4}, & \\
& \nonumber\mathcal{B}(J/\psi\to K^{+}K^{-})=[2.86\pm 0.09(\mathrm{stat})\pm 0.19(\mathrm{syst})]\times 10^{-4}, & \\
& \mathcal{B}(J/\psi\to K_{S}K_{L})=[2.62\pm 0.15(\mathrm{stat})\pm 0.14(\mathrm{syst})]\times 10^{-4}. &
\end{eqnarray}  
These are listed in Table~\ref{tbl:brresult}, and in Table~\ref{tbl:deltaresult} with the statistical and systematic errors 
combined in quadrature.
The above branching fractions for $J/\psi\to \pi^{+}\pi^{-}$ and $K^{+}K^{-}$ are
consistent with those reported by Mark III~\cite{spear}, and have factors two and three 
smaller uncertainties, respectively. Our branching fraction for $J/\psi \to K_{S}K_{L}$ 
based on $334\pm19$ well resolved events, as shown in Fig.~4~(f), is a factor 2.6 larger than 
$\mathcal{B}(J/\psi\to K_{S}K_{L})=(1.01\pm 0.18)\times 10^{-4}$ reported by 
Mark III with 74 identified counts, obtained by ``stringent cuts" to remove 
$J/\psi \to \rho^{0}\pi^{0}$ and $J/\psi \to K_{S}\bar{K}^{*}(898)$ decays, 
and is $\sim 40 \%$ larger than 
$\mathcal{B}(J/\psi\to K_{S}K_{L})=(1.82\pm 0.13)\times 10^{-4}$ reported by 
BES II~\cite{bes} with $2155\pm 45$ identified events. In the large statistics BES II
measurements the peak due to $J/\psi \to \overline{K}^{*0}(892)K_{S}$, 
$\overline{K}^{*0} \to K_{L}\pi^{0}$, $K_S\to\pi^-\pi^+$, overlapped with the direct $J/\psi \to K_{S}K_{L}$,
$K_{S}\to \pi^{+}\pi^{-}$ peak,
and strong cuts had to be made to resolve the two peaks. As shown in Figs.~4~(c,f), in our
measurements the two peaks are completely resolved.
Further, MC calculations confirm that with our selections the decay  $J/\psi \to \pi^{0}\rho^{0}$, $\rho^{0} \to \pi^{+}\pi^{-}$ also does not make any contribution
under the $J/\psi \to K_{S}K_{L}$, $K_{S}\to \pi^{+}\pi^{-}$ peak at $X(K_{S})=1.0$.

As for $\psi(2S)$, we calculate $\delta(J/\psi)$ using Eq.~(1), and obtain $\delta(J/\psi)=(73.6\pm5.6)^\circ$, as compared to $(88\pm11)^\circ$ obtained using the branching fractions measured by Mark~III.  These are listed in Table~\ref{tbl:deltaresult}.

\begin{figure*}[tb!]         
\begin{center}         
\includegraphics[width=\textwidth]{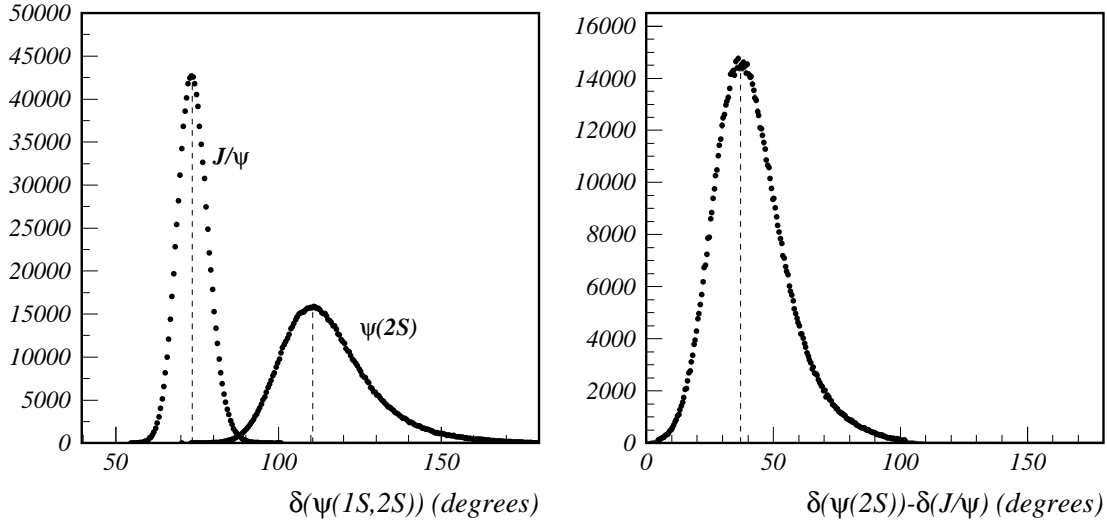}   
\end{center}         
\caption{(Left) The relative phase angle distributions $\delta(\psi)$ for $\psi(2S)$ and $J/\psi$ obtained from toy MC simulations.  The results are $\delta(\psi(2S)) = (110.5^{+16.0}_{-9.5})^\circ$ and $\delta(J/\psi)=(73.5^{+5.0}_{-4.5})^\circ$.  (Right) Toy MC distribution of the difference of relative phase angles for $\psi(2S)$ and $J/\psi$, $\Delta \delta = \delta(\psi(2S))-\delta(J/\psi)=(37.0^{+16.5}_{-10.5})^\circ$.}
\end{figure*}

\section{Monte Carlo Based Evaluation of $\delta(\psi)$ and the Difference $\Delta\delta \equiv \delta(\psi(2S))-\delta(J/\psi)$}

The individual measured branching fraction values have distributions which are conventionally stated in terms of $1\sigma$ errors, as listed in Table~I. Using only the central values to evaluate $\delta(\psi)$ according to Eq.~(1) is not correct.  The more correct procedure is to make Monte Carlo evaluation of Eq.~(1) taking account of random associations of the branching fraction values in the distributions for the three decays.  We have made such toy MC evaluations of both $\delta(\psi(2S))$ and $\delta(J/\psi)$.  As shown in Fig.~5~(left), the large error ($\pm30\%$) in $\psi(2S)\to\pi^+\pi^-$ branching fraction leads to a very asymmetric MC  distribution for $\delta(\psi(2S))$, whereas the much smaller error ($\pm12\%$) for $J/\psi\to\pi^+\pi^-$ results in a much smaller asymmetry in the distribution for $\delta(J/\psi)$.  If we adopt the usual definition of the $1\sigma$ error as that which includes 68\% of the area on each side of the peak of a distribution, our results are
\begin{eqnarray}
& \delta(\psi(2S))_{PP} = (110.5^{+16.0}_{-9.5})^\circ,~~\mathrm{and}~~ \delta(J/\psi)_{PP}=(73.5^{+5.0}_{-4.5})^\circ. &
\end{eqnarray}
The difference, whose MC distribution is illustrated in Fig.~5~(right), is 
\begin{equation}
\Delta\delta\equiv\delta(\psi(2S))_{PP}-\delta(J/\psi)_{PP}=(37.0^{+16.5}_{-10.5})^\circ.
\end{equation}
We consider the above estimates of $\delta(\psi(2S))_{PP}$, $\delta(J/\psi)_{PP}$, and their difference to be our final results.

\section{Conclusions}

We have made large statistics measurements of the branching fractions for the decays of $\psi(2S)$ and $J/\psi$ to pseudoscalar pairs $\pi^+\pi^-$, $K^+K^-$, and $K_SK_L$.  Our branching fraction results have errors which are factors two to five smaller than the previously published results. Using these branching fractions we have made calculations of the phase angle differences between the electromagnetic and strong decays for both $\psi(2S)$ and $J/\psi$,  taking proper account of the distributions of the branching fraction values.  Our results are nearly a factor two more precise than the previously published results.

\begin{acknowledgments}
This measurement was done using CLEO data, and as members of the former CLEO
Collaboration we thank it for this privilege.
We wish to thank Jon Rosner for his helpful comments and suggestions.
This research was supported by the U.S. Department of Energy and the National
Science Foundation.
\end{acknowledgments}


\begin{thebibliography}{99}
\bibitem{explain} The pQCD prediction of the constancy of $Q^{2}F(Q^{2})$ for vector mesons,
and the corresponding prediction for protons, have been experimentally 
confirmed~\cite{zweber}. For example, for pions $Q^{2}F(\pi)=0.94\pm 0.06$ GeV$^{2}$ at
$Q^{2}=9.6$ GeV$^{2}$, and $1.01\pm 0.13$ GeV$^{2}$ at $Q^{2}=13.48$ GeV$^{2}$~\cite{zweber}.
\end{thebibliography}
\end{document}